\documentclass[twocolumn,amssymb,noeprint,floatfix,superscriptaddress,aps, prl, reprint]{revtex4-2}

\usepackage{float}
\usepackage{changes}
\usepackage[utf8]{inputenc}
\usepackage{blindtext}
\usepackage{chemformula}
\usepackage[T1]{fontenc}
\usepackage{graphicx}
\usepackage{natbib}
\usepackage{textgreek}
\usepackage{siunitx}
\usepackage[utf8]{inputenc}
\usepackage{amsmath}
\usepackage{booktabs} 
\usepackage{xcolor}
\usepackage{xr}
\definecolor{hypercolor}{RGB}{181, 51, 137}
\usepackage[linkcolor=hypercolor, colorlinks=true, citecolor=hypercolor, urlcolor=hypercolor]{hyperref}
\usepackage{cleveref}
\usepackage[version=4]{mhchem}
\usepackage{bm}
\newcommand{\SIerr}[3]{\SI[separate-uncertainty = true]{#1(#2)}{#3}}
\newcommand{\SIerrBold}[3]{%
  \ensuremath{%
    \mathbf{#1} \boldsymbol{\pm} \mathbf{#2}~%
  }%
  \textbf{\si{#3}}%
}
\newcommand{\SIBold}[2]{%
  \ensuremath{\mathbf{#1}~}%
  \textbf{\si{#2}}%
}

\date{\today}

\newcommand{\Tbath}{$T_\text{bath}$}
\newcommand{\Pexc}{$P_\text{exc}$}
\newcommand{\Eexc}{$E_\text{exc}$}

\newcommand{\gzwo}[1][\empty]{%
  g^{(2)}(\tau%
  \ifx#1\empty
  \else
    \!=\!0%
  \fi
  )%
}

\begin{document}
\newcommand{\wsi}{Walter Schottky Institut and Physics Department, Technical University of Munich, Am Coulombwall 4a, 85748 Garching, Germany.}
\newcommand{\mcqst}{MCQST, Schellingstrasse 4, 80799 München, Germany.}
\newcommand{\kth}{KTH Royal Institute of Technology, Department of Applied Physics, Albanova University Centre, Roslagstullsbacken 21, 106 91 Stockholm, Sweden}
\newcommand{\mich}{Department of Electrical Engineering and Computer Science, University of Michigan, 1301 Beal Avenue, Ann Arbor, Michigan 48109, United States}
\newcommand{\weizmann}{Department of Molecular Chemistry and Materials Science, Weizmann Institute of Science, 7610001 Rehovot, Israel.}
\newcommand{\Duisburg}{Faculty of Physics, University of Duisburg-Essen, Germany.}
\newcommand{\takashi}{Research Center for Materials Nanoarchitectonics, National Institute for Materials Science,  1-1 Namiki, Tsukuba 305-0044, Japan}
\newcommand{\watanabe}{Research Center for Electronic and Optical Materials, National Institute for Materials Science, 1-1 Namiki, Tsukuba 305-0044, Japan}

\author{Katja Barthelmi}
\affiliation{\wsi}
\affiliation{\mcqst}

\author{Tomer Amit}
\affiliation{\weizmann}

\author{Mirco Troue}
\affiliation{\wsi}
\affiliation{\mcqst}

\author{Lukas Sigl}
\affiliation{\wsi}

\author{Alexander Musta}
\affiliation{\wsi}

\author{Tim Duka}
\affiliation{\wsi}

\author{Samuel Gyger}
\affiliation{\kth}

\author{Val Zwiller}
\affiliation{\kth}

\author{Matthias Florian}
\affiliation{\mich}

\author{Michael Lorke}
\affiliation{\Duisburg}

\author{Takashi Taniguchi }
\affiliation{\takashi}

\author{Kenji Watanabe}
\affiliation{\watanabe}

\author{Christoph Kastl}
\affiliation{\wsi}
\affiliation{\mcqst}

\author{Jonathan Finley}
\affiliation{\wsi}
\affiliation{\mcqst}

\author{Sivan Refaely-Abramson}
\affiliation{\weizmann}

\author{Alexander Holleitner}
\email[]{holleitner@wsi.tum.de}
\affiliation{\wsi}
\affiliation{\mcqst}

\begin{abstract}
We explore the zero-phonon line of single photon emitters in helium-ion treated monolayer \ce{MoS2}, which are currently understood in terms of single sulfur-site vacancies. By comparing the linewidths of the zero-phonon line as extracted directly from optical spectra with values inferred from the first-order autocorrelation function of the photoluminescence, we quantify bounds of the homogeneous broadening and of phonon-assisted contributions. The results are discussed in terms of both the independent boson model and ab-initio results as computed from GW and Bethe-Salpeter equation approximations. 

\end{abstract}

\title{Zero-phonon line emission of single photon emitters in helium-ion treated \texorpdfstring{\ce{MoS2}}{MoS2}}
\date{December 1, 2025}
\maketitle

\section{1. Introduction}

Single-photon sources are crucial components for the advancement of quantum technologies, including quantum communication, computation, simulation, and precision measurements~\cite{aharonovich_solid-state_2016, Rodt2021, esmann_solid-state_2024}. Recently, single-photon emitters (SPEs) have been identified in both bulk and two-dimensional (2D) van der Waals (vdW) materials~\cite{Tonndorf2015, Koperski2015,Srivastava2015,He2015,Chakraborty2015, Tran2016quantum, Montblanch2023}. As a result, these materials are gaining increasing attention as a novel solid-state platform particularly for quantum photonics~\cite{raj_single_2025}, providing an alternative to more established systems like semiconductor quantum dots~\cite{Hepp2019} or diamond color centers~\cite{Atature2018}. Several key features of SPEs have been demonstrated for such vdW-based material platforms, including room-temperature operation~\cite{Tran2016quantum}, high quantum efficiencies~\cite{nikolay_direct_2019}, gate tunability \cite{chakraborty_voltage-controlled_2015, Branny2016, hotger_gate-switchable_2021}, and position accuracy \cite{Mitterreiter2020}, supporting their prospective integration into photonic and optoelectronic circuits~\cite{aharonovich_solid-state_2016, Rodt2021, esmann_solid-state_2024, Montblanch2023, raj_single_2025, Atature2018, carbone_creation_2025-1}. 

The optical linewidth of the SPEs depends strongly on the underlying vdW material platforms, the employed fabrication methods, as well as on the excitation and measurement schemes. For strain-engineered SPEs within \ce{WSe2} monolayers, the linewidths can be on the order of \SI{120}{\micro\eV} ($\sim$ \SI{30}{G\Hz}) \cite{Koperski2015,Srivastava2015, He2015,Chakraborty2015, Palacios-Berraquero2017} and as small as \SI{75}{\micro\eV} ($\sim$ \SI{18}{G\Hz}) for SPEs in monolayers encapsulated in hBN~\cite{Parto2021}. In contrast, impurity-bound and strain-induced SPEs in \ce{WS2} tend to have broader linewidths, often reaching several hundreds of \SI{}{\micro\eV}~\cite{Loh2024,Cianci2023}, while SPEs in strained \ce{MoSe2} can exhibit linewidths on the order of \SI{150}{\micro\eV} ($\sim$ \SI{36}{G\Hz}). 
Substantially larger linewidths, on the order of \SI{}{\milli\eV}, have been reported for SPEs in pre-strained MoTe$_2$, InSe, and GaSe, likely attributable to a pronounced spectral diffusion~\cite{Tonndorf2017, Zhao2021, Zhao2025}. In \ce{MoS2}, SPEs have demonstrated linewidths of several hundreds of \SI{}{\micro\eV} so far~\cite{Klein2019, Mitterreiter2021, Wang2022, Dash2025}. In most cases, SPEs in vdW materials exhibit a strong phonon coupling ~\cite{MichaelisdeVasconcellos2022}, which manifests in the emission spectrum as a zero-phonon line superimposed by a pronounced phonon sideband, thereby contributing to additional spectral broadening. 
The influence of resonant excitation schemes~\cite{kumar_resonant_2016-1,Dietrich2018} is particularly evident for SPEs in hBN~\cite{Tran2016quantum}: under non-resonant excitation, linewidths are typically in the \SI{}{\milli\eV} range, whereas a resonant excitation enables the observation of Fourier-limited single-photon emission with linewidths as narrow as \SI{0.2}{\micro\eV} ($\sim$ \SI{50}{M\Hz})~\cite{Dietrich2018}. 

Here, we discuss the photoluminescence (PL) of SPEs in helium-ion treated \ce{MoS2} \cite{Klein2019, KleinSigl2021}, which can be understood in the framework of single sulfur-site vacancies~\cite{Mitterreiter2020, Mitterreiter2021, hotger_photovoltage_2023, Hotger2023}. We examine the spectral lineshape of the luminescence to get insights into the broadening of the zero-phonon line (ZPL) and the underlying exciton-phonon interactions as a function of the experimental bath temperature and the applied excitation power by utilizing the independent boson model (IBM).
Based on the IBM, we estimate an upper limit of the linewidth of the ZPL of SPEs in \ce{MoS2} encapsulated in hBN to be on the order of \SI{110}{\micro\eV} ($\sim$ \SI{27}{G\Hz}) for a bath temperature below \SI{20}{\K}, and based on the asymptotic behavior of the high energy tail of the ZPL, we estimate a lower bound on the order of 30 - \SI{60}{\micro eV} ($\sim$ 7 - 10 GHz) for a bath temperature below \SI{2}{\K}. Moreover, we compare the derived values for both the zero-phonon line and the phonon-related photon emission processes with values derived from complementary measurements on the first-order autocorrelation function of the luminescence. We find good agreement between the different experimental methods. The inferred timescales of 6 - \SI{22}{\pico\s} are consistently within the range of lifetimes for defect excitons predicted by many-body perturbation theory within the GW approximation and Bethe-Salpeter equation approach (GW-BSE)~\cite{Amit2022}. We contrast these results with lifetimes as measured in time-resolved photoluminescence experiments demonstrating a single exponential decay time of several tens of nanoseconds at low excitation intensity and bath temperature. Our results reveal detailed microscopic insights into the emission process of single photons in \ce{MoS2}-based SPEs, opening the possibility to utilize them in photonic and optoelectronic applications~\cite{aharonovich_solid-state_2016, Rodt2021, esmann_solid-state_2024, Montblanch2023, raj_single_2025, Atature2018, carbone_creation_2025-1}.

\begin{figure}[tp]
\centering
\includegraphics[scale=1]{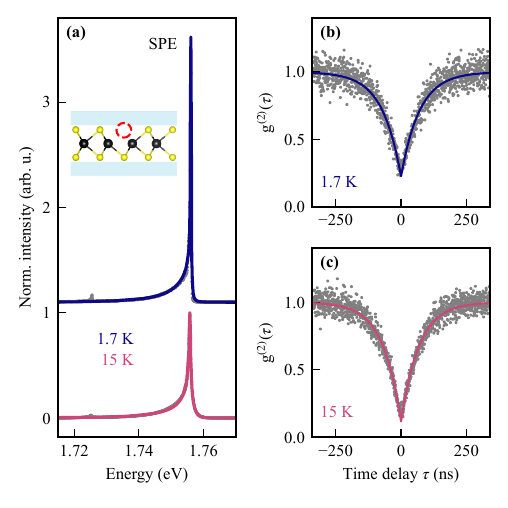}
\caption{\textbf{Defect PL emission and second-order autocorrelation measurement.} \textbf{(a)} Photoluminescence spectra of a single-photon emitter (SPE) in helium-ion treated monolayer \ce{MoS2} at \Tbath~$=\SI{1.7}{\K}$ (blue) and \SI{15}{\K} (purple). The data normalized to the maximum for \Tbath~$=\SI{15}{\K}$. Experimental parameters are \Pexc~$=\SI{500}{\nano\W}$, \Eexc~$=\SI{1.94}{\eV}$. The data (gray) are fitted by an independent boson model (solid lines) accounting for coupling of the emitter to acoustic phonons. Inset: sketch of the hBN/\ce{MoS2}/hBN sample structure with a sulfur-site defect highlighted by a red dashed circle. \textbf{(b)} and \textbf{(c)}: Measured second-order autocorrelation function $g^{(2)}(\tau)$ (gray dots) of the defect emission with corresponding fits (solid lines) for \Tbath~$=\SI{1.7}{\K}$ (b) and $\SI{15}{\K}$ (c)}
\label{fig: 1}
\end{figure}
\section{2. Experimental details and results}
\subsection{2.1 Sample design and characterization}
The samples consist of monolayers of \ce{MoS2} encapsulated by a top and a bottom layer of hBN [cf. inset of Fig \ref{fig: 1}(a)]. The heterostructures are positioned on top of \ce{SiO2}/Si-substrates (\SI{295}{\nano\m}/\SI{525}{\micro\m}) and subsequently irradiated with a focused He-ion beam, such that SPEs are generated at certain positions (details as in Refs.~\cite{Barthelmi2020,KleinSigl2021,Hotger2023, barthelmi_spectrally_2025}). Fig.~\ref{fig: 1}(a) shows the photoluminescence (PL) spectra of one of the SPEs at a bath temperature of \Tbath~$=\SI{1.7}{\K}$ and \SI{15}{\K} excited at a photon energy of \Eexc~$=\SI{1.94}{\eV}$ (cf. Supporting Information) and an excitation power of \Pexc~$=\SI{500}{\nano\W}$ at a spot size of one micrometer. The PL is detected with the help of a spectrometer and a charge-coupled device. For both temperatures, the data are well fitted by the IBM (blue and purple lines), as will be discussed in detail below. Figures~\ref{fig: 1}(b) and (c) present the second-order autocorrelation function $g^{(2)}(\tau)$ of the luminescence as measured at \Tbath~$=\SI{1.7}{\K}$ and \SI{15}{\K}, with $\tau$ the time delay between two detected photons. For the autocorrelation measurements, we utilize a spectral band pass ($\pm 13$ nm) filter with a central wavelength of ($711$ nm) ($\sim$ \SI{1.744}{\eV}). A fit gives $g^{(2)}(0)$~= $0.23~\pm~ 0.01$ \SIerr{0.12}{1}{} for the data at \SI{1.7}{\K} (\SI{15}{\K}) [lines in Figs.~\ref{fig: 1}(b) and (c)]. In both cases, the value of $g^{(2)}(0)$ is below 0.5, which is indicative that the investigated emitter is an SPE at both temperatures. Moreover, the fits give a characteristic timescale for the single photon decay of \SIerr{71}{2}{\nano\s} at \SI{1.7}{\K} and \SIerr{68}{1}{\nano\s} at \SI{15}{\K}.\\

\begin{figure}[t!]
\centering
\includegraphics[scale=1]{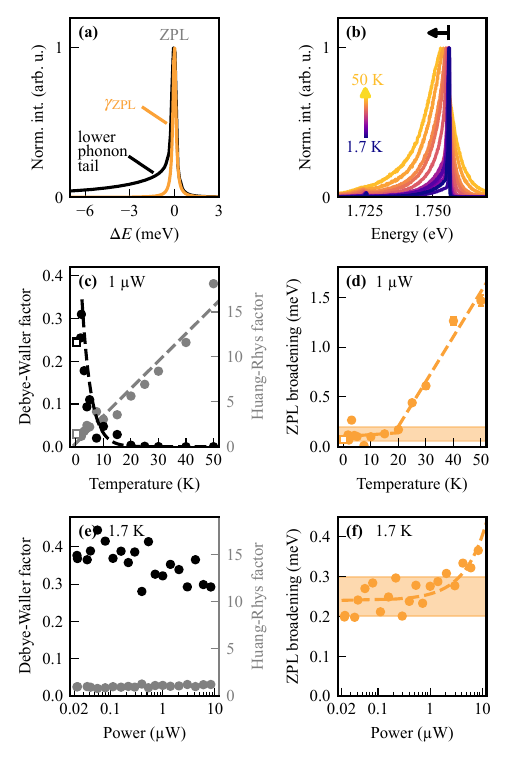}
\caption{\textbf{Temperature and excitation power dependence of SPE-related photoluminescence.} \textbf{(a)} Sketch of the IBM highlighting the contribution from the ZPL (yellow) and a lower tail related to emission of acoustic phonons. \textbf{(b)} Normalized PL spectra for \Tbath~ranging from~$\SI{1.7}{\K}$ (blue) to \Tbath~$=\SI{50}{\K}$ (yellow) with \Eexc~$=\SI{1.94}{\eV}$ and \Pexc~$=\SI{500}{\nano\W}$. Arrow on top highlights the red-shift of the maximum for an increasing \Tbath. \textbf{(c)} Extracted Debye-Waller factor (black dots, DWF) and Huang-Rhys factor (gray dots, HRF) vs. \Tbath~for \Pexc~$=\SI{1}{\micro\W}$. Dashed lines are guides to the eye. \textbf{(d)} ZPL broadening (yellow dots) vs. \Tbath. The open square represents data from a second sample with \Tbath~below $\SI{200}{\milli \K}$. Dashed lines highlight the different slopes below and above \SI{20}{\K}. Yellow area highlights the data spread at low temperature for this cool-down. \textbf{(e)} Excitation power dependence of the extracted Debye-Waller (black) and Huang-Rhys (gray) factor for \Tbath~$=\SI{1.7}{\K}$. \textbf{(f)} ZPL broadening vs. excitation power (\Pexc). Yellow area highlights the data spread for low laser powers during this second cool-down. Dashed line highlights a linear trend.}
\label{fig: 2}
\end{figure}
\subsection{2.2 Photoluminescence Spectra}

Figure \ref{fig: 2}(a) sketches the spectral contributions in the framework of an IBM, as it is typically applied for a simplified description of the luminescence spectra of SPEs in various van der Waals materials, including \ce{MoS2}~\cite{Klein2019, MichaelisdeVasconcellos2022}, \ce{WSe2}~\cite{vannucci2024single}, or hBN~\cite{khatri2019phonon}. At low temperatures, where phonon absorption is negligible, the contributions of the model can be decomposed a Lorentzian (yellow line) at the emission energy of the ZPL (i.e. $\Delta E = 0$) accounting for dephasing and radiative broadening and a red-shifted luminescence tail capturing emission of acoustic phonons. The overall strength of the phonon coupling is often summarized conveniently in terms of the Huang-Rhys factor (HRF) $S$, which can be understood by an average number of phonons coupled into the optical transition weighted by phonon energy and density of states~\cite{Huang1950,Zhang2019,Cochrane2021}. 
Figure~\ref{fig: 2}(b) presents the normalized experimental PL spectra for the emitter as in Fig.~\ref{fig: 1} for \Tbath~ranging from \SI{1.7}{\K} (blue) to \SI{50}{\K} (yellow). We note that the amplitude of the luminescence of the investigated SPEs decreases with temperature [cf. Fig.~\ref{fig: 1}(a)]~\cite{Klein2019}. The chosen normalization, however, visualizes both the overall broadening of the luminescence and the redshift of the ZPL (top arrow) with increasing temperature (cf. Supporting Information for original spectra).

Each of the luminescence spectra is fitted by applying the IBM as sketched in Figure~\ref{fig: 2}(a). Figure~\ref{fig: 2}(c) displays the corresponding values of the temperature-dependent HRF $S(T)$ as a function of \Tbath\space (gray data), which explicitly includes the phonon occupation. With increasing \Tbath, the temperature-dependent HRF increases monotonically; i.e., in this picture, the average number of phonons involved per optical transition increases with temperature because phonon emission and absorption processes scale with $n_\mathrm{ph}(\mathbf{q}) +1$ and $n_\mathrm{ph}(\mathbf{q})$, respectively, with $n_\mathrm{ph}(\mathbf{q})$ being the phonon occupation number at wave vector $\mathbf{q}$. A linear fit of the scaling at $T_\mathrm{bath}>\SI{10}{K}$ yields a slope of \SI{0.3}{\per\K}. Below about \SI{5}{K}, the effective HRF approximates to good precision the zero-temperature HRF, where we find an average of \SIerr{1.6}{3}{}. From the HRF, one can derive the corresponding temperature-dependent Debye-Waller factor (DWF)~\cite{Zemla2020}:
\begin{equation}
\label{eq: DWF}
    \text{DWF} = e^{-S(T)},    
\end{equation}
which describes the ratio between the luminescence intensity of the ZPL with respect to the total emission~\cite{Zemla2020,Vuong2016,Tran2016quantum}. Fig.~\ref{fig: 2}(c) depicts the temperature-dependent DWF as calculated from eq.~\ref{eq: DWF} for each \Tbath\space by black dots, while the black dashed line represents the values, when eq.~\ref{eq: DWF} is applied to the linear fit of the HRF. We deduce that for the particular set of measurements, up to $\sim$\SI{30}{\%} of the luminescence is emitted via the ZPL at low temperatures, and that the fraction decreases exponentially for increasing temperature. For comparison, the open squares highlight the results from an SPE on a second sample with a maximum of $\sim$\SI{25}{\%} (black) and a consistent HRF (gray) as measured in a second set-up at \Tbath~$\sim\SI{200}{\milli\K}$. Consistent with the freezing out of phonon absorption processes at low-temperature, the linewidth of the ZPL $\gamma_\text{ZPL}$ as fitted wih the IBM increases only weakly with at \Tbath $< \SI{20}{k}$ by about \SIerr{1.9}{1}{\micro\eV\per\K} and an intercept of \SIerr{102}{1}{\micro\eV} at \SI{0}{\K} [Fig.~\ref{fig: 2}(d)]. Again, the open square depicts the corresponding value (\SI{70.4}{\micro\eV}) at \Tbath~$\sim\SI{200}{\milli\K}$ for the SPE on the second sample. 
Above \SI{20}{\K}, the spectra in Fig.~\ref{fig: 2}(b) broaden also for an emission energy above the ZPL. In the IBM-fit, $\gamma_\text{ZPL}$ shows a corresponding increase with about \SI{45}{\micro\eV\per\K} [cf. (Fig.~\ref{fig: 2}(d)]. We tentatively explain this broadening of the spectra by carrier-phonon interaction processes, as discussed below.

\begin{figure}
\centering
\includegraphics[scale=1]{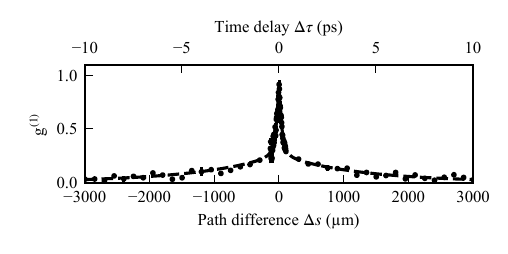}
\caption{\textbf{First-order auto-correlation measurement of the SPE.} $g^{(1)}$ of the SPE as a function of the path difference $\Delta s$ (bottom) of the utilized interferometer and correspondingly, the time delay $\Delta \tau$ (top) of the interfering signals. Data are fitted by the sum of a Lorentzian and two exponential tails (dashed lines). Experimental parameters are \Tbath~$=\SI{1.7}{\K}$ , \Eexc~$=\SI{1.94}{\eV}$ and \Pexc~$=\SI{250}{\nano\W}$. 
}
\label{fig: 3}
\end{figure}

After a temperature-cycle to room temperature, we measure the PL spectra of the same SPE as a function of the laser power at \Tbath = \SI{1.7}{\K}. Figs.~\ref{fig: 2}(e) and (f) show the results of the analysis as in Fig.~\ref{fig: 2}(a). The deduced DWF now reaches values of about 40{\%} at the lowest investigated powers for this set of experiments [left $y$-axis in Fig.~\ref{fig: 2}(e)] decreasing slightly with increasing excitation power. Correspondingly, the fitted value of HRF increases from \SI{1.15} (below \SI{1}{\micro\W}) with a linear slope of \SI{0.03}{\per\micro\W} (right axis). The fitted value of $\gamma_\text{ZPL}$ increases linearly with excitation power starting from \SI{240}{\micro\eV} with \SI{16}{\micro\eV\per\micro\W} [Fig.~\ref{fig: 2}(f)]. Generally, the power dependence of the DWF, the HRF, and $\gamma_\text{ZPL}$ can be understood by the impact of local, laser-induced heating.

\subsection{2.3 Temporal coherence}

To further characterize the ZPL linewidth as deducted from the same SPE over multiple cooldown cycles, we perform first-order autocorrelation measurements on the SPE. Figure~\ref{fig: 3} depicts $g^{(1)}$ at \Tbath $=\SI{1.7}{\K}$ with a distinct maximum close to zero path difference $\Delta s$ and corresponding time difference $\Delta \tau$ of the utilized interferometer as well as a further decaying contribution for larger values of $\Delta$s and $\Delta \tau$. We fit the center contribution with a Lorentzian function with a FWHM of \SIerr{280}{10}{\femto\s} (cf. Supporting Information,) which corresponds to an energetic linewidth of \SIerr{2.336}{2}{\milli\eV} (cf. Table ~\ref{tab: 2}). We interpret the values to resemble the phonon-related PL of the SPE [cf. Fig.~\ref{fig: 2}(a)] and its corresponding short temporal coherence time. We further fit the decaying contribution of $g^{(1)}$ at long time scales with an exponential function and derive a decay time of $\tau _{g^{(1)}}$ = \SIerr{6.0}{3}{\pico\s} (cf. Supporting Information), which we interpret to be the temporal coherence time of the ZPL of the SPE. We note that this interpretation is consistent with the Wiener-Khinchin theorem, since an exponential decay in the time domain is Fourier-transformed to a Lorentzian function in the energy domain, as it is applied to describe the ZPL within the above discussed IBM [cf. Fig.~\ref{fig: 2}(a)]. The corresponding energetic linewidth is $\gamma_{g^{(1)}}$ = \SIerr{110}{6}{\micro\eV}, which is consistent with the lowest value of the homogeneous linewidth at \Tbath = \SI{1.7}{\K} as derived with the help of the IBM in Fig.~\ref{fig: 2}. For the $g^{(1)}$-measurement, we note that \Pexc$=\SI{250}{\nano\W}$ is well within the low-power regime where heating effects are still negligible [cf. Fig. ~\ref{fig: 2}(f)]. \\

\begin{figure}[ht!]
\centering
\includegraphics[scale=1]{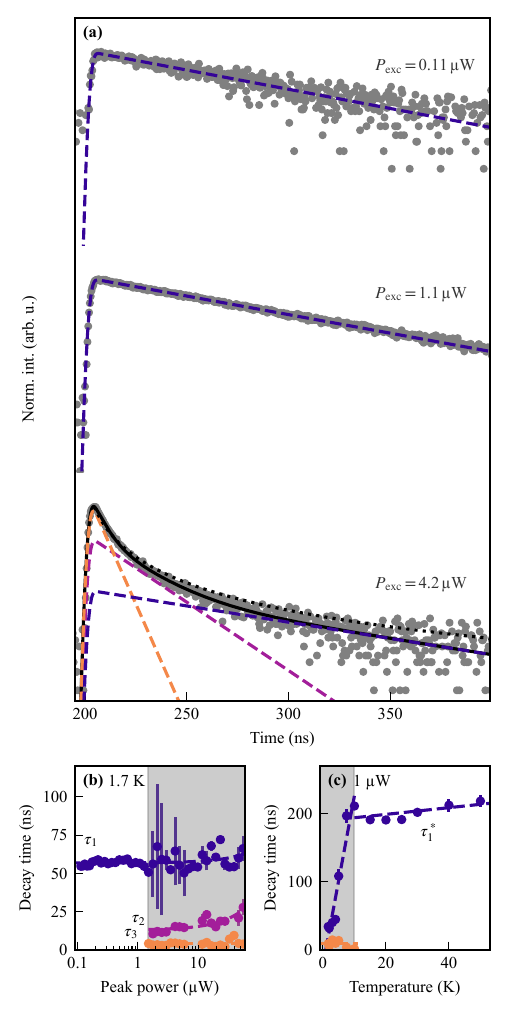}
\caption{\textbf{Time-resolved photoluminescence of the SPE.} \textbf{(a)} Data are measured at \Tbath~$=\SI{1.7}{\K}$ and displayed in gray with exponential fits in blue, purple, and orange. Black line at bottom panel is sum of blue, purple, and orange. Black dotted line is a polynomial fit (cf. Supporting Information). \Pexc~as indicated. \textbf{(b)} Power dependence of the extracted decay times $\tau_1$, $\tau_2$, and $\tau_3$. Colors as in (a). \textbf{(c)} Temperature dependence of the extracted decay times at \Eexc~$=\SI{1.94}{\eV}$ with dashed lines as guides to the eye. The extracted decay time at high \Tbath~ is baptized $\tau^*_1$.}
\label{fig: 4}
\end{figure}

\subsection{2.4 Time-resolved photoluminescence}

\begin{figure*}[t!]
\centering
\includegraphics[scale=1]{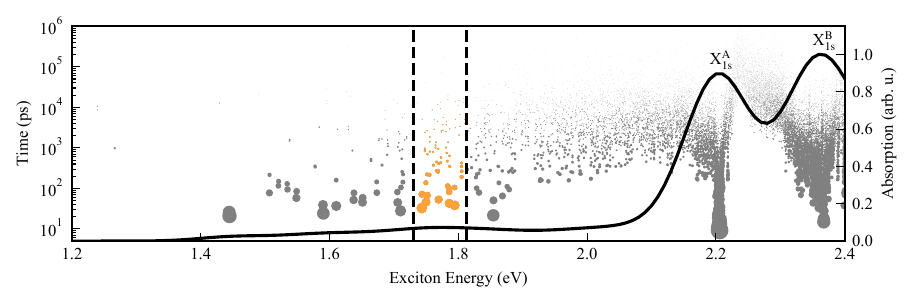}
\caption{\textbf{Ab initio results of radiative decay times for a sulfur vacancy in \ce{MoS2}.} The black line displays the GW-BSE computed absorption spectrum as a function of energy, while the calculated radiative decay times for different exciton states in \ce{MoS2} are displayed by dots, with corresponding oscillator strength indicated by the dot size. The colored dots around \SI{1.75}{\eV} highlight the energy comparable to the photon emission of the SPE as in the rest of the manuscript. }
\label{fig: 5}
\end{figure*}

In a next step, we focus on the time-resolved photoluminescence of the SPE as in Fig.~\ref{fig: 1}. For this experiment, the sample is excited with a pulsed laser with an excitation energy of \Eexc~$=\SI{1.94}{\eV}$ at a repetition frequency of \SI{2}{\mega\Hz} and a pulse duration of \SI{2}{\nano\s}. The emission is detected with an avalanche photo diode (APD). As known for SPEs in \ce{MoS2}, the SPE-related photoluminescence exhibits the highest intensity when the excitation exploits the oscillator strength of the 1s-exciton X$^A_\text{1s}$ or the one of possible trion-states of \ce{MoS2} (cf. Supporting Information). To the best of our knowledge, a resonant excitation of the SPEs in helium-ion treated \ce{MoS2} has not been achieved yet. Fig.~\ref{fig: 4}(a) shows the results for three excitation powers \Pexc = \SI{0.11}{\micro\W}, \SI{1.1}{\micro\W}, and \SI{4.2}{\micro\W} at \Tbath $=\SI{1.7}{\K}$. The luminescence is fitted by exponential decay curves. We observe that the data at low excitation powers can be described very consistently by one exponential decay time, which we call $\tau_1$ (blue dashed line in Fig.~\ref{fig: 4}(a) for \Pexc = \SI{0.11}{\micro\W} and \SI{1.1}{\micro\W}), while at higher \Pexc, this single-exponential description seems to be insufficient to grasp the dynamics of the SPE-related photoluminescence (blue dashed line in \ref{fig: 4}(a) for \Pexc = \SI{4.2}{\micro\W}). Phenomenologically, we include two further exponential decays with characteristic times $\tau_2$ (purple dashed line) and $\tau_3$ (orange) in order to sufficiently describe the data within the given noise amplitude (black line as the sum of all three exponential functions). The dotted black line is a polynomial fit to the data as discussed below and in the Supporting Information. Fig.~\ref{fig: 4}(b) shows the power dependence of the extracted exponential decay times at a constant bath temperature (\Tbath~= \SI{1.7}{\K}). The decay time $\tau_1$ (blue) has a mean value of \SIerr{58}{1}{\nano\s} in the regime with a mono-exponential decay. In the case of a non-mono-exponential decay [shaded area in Fig. \ref{fig: 4}(b)], it exhibits a small linear slope of \SI{0.1}{\nano\s\per\micro\W}. The time $\tau_2$ (purple) has a value of \SIerr{11}{2}{\nano\s} at the lowest \Pexc~where it occurs, and it increases linearly with a slope of \SI{0.2}{\nano\s\per\micro\W}. The shortest time $\tau_3$ (orange) seems to be independent from the excitation power and it has a mean value of \SIerr{4.1}{4}{\nano\s}. We note that above \SI{56}{\micro\W}, the SPE's emission becomes unstable for this particular experiment.
Fig.~\ref{fig: 4}(c) depicts the temperature dependence of the extracted decay times at $ \SI{1}{\micro W}$. Interestingly, at a higher temperature (above \SI{10}{\K}, the SPE exhibits again a regime with a mono-exponential decay. The corresponding decay time $\tau^*_1$ has a value on the order of approx. \SI{200}{\nano\s} (cf. Supporting Information). The transitions from mono- to non-mono-exponential decay dynamics as a function of excitation power and bath temperature are discussed later below.

\section{3. Ab-initio simulation}

Next, we perform a GW-BSE calculation, yielding excitonic states for a sulfur vacancy in \ce{MoS2} as detailed in refs. \cite{Refaely-Abramson2018, Amit2022}. Each state is denoted as $S$, with corresponding energy $E_\text{S}$ and oscillator strength $f_\text{S}$. We calculate the first principles radiative decay time for each many-body excitonic state, $\tau_\text{S}$, as given in atomic units by: $\tau_\text{S}^{-1}=\frac{4\pi}{cA} E_\text{S} \left|f_\text{S}\right|^2 $, with $c$ the speed of light and $A$ the calculated unit cell area. Figure~\ref{fig: 5} displays the GW-BSE computed results for a monolayer \ce{MoS2} 5x5 supercell with a single sulfur vacancy (see Supporting Information for full computational details). Figure~\ref{fig: 5} shows the calculated absorption spectrum (black line) as well as the calculated radiative decay times of excitonic states (dots) as a function of state energy, with dot sizes corresponding to the oscillator strengths. The dashed lines highlight the energy region around \SI{1.75}{\eV}, which is comparable to the SPE's emission energy of the ZPL as in Fig.~\ref{fig: 1}(a). The shortest-lived computed exciton states in this energy range exhibit lifetimes of a few tens of \SI{}{\pico\s}.

\begin{table}[h!]
  \centering
    \begin{tabular}{l|r|r}
    & linewidth & time\\
\toprule
IBM& \SIerrBold{101}{1}{\micro\eV}& \SIerr{6.5}{1}{\pico\s}\\
$g^{(1)}(\tau)$& \SIerr{110}{6}{\micro\eV}& \SIerrBold{6.0}{0.3}{\pico\s}\\
& \SIerr{2.35}{8}{\milli\eV} & \SIerrBold{280}{10}{\femto\s}\\
\textbf{$\tau_1$} & \SIerr{12.0}{2}{\nano\eV}& \SIerrBold{55}{1}{\nano\s}\\
\textbf{$\tau_2$} & \SIerr{60}{11}{\nano\eV} & \SIerrBold{11}{2}{\nano\s}\\
\textbf{$\tau_3$} & \SIerr{199}{6}{\nano\eV}&\SIerrBold{3.3}{0.1}{\nano\s}\\
$\tau_1^*$ (high \Tbath)& $\sim$\SI{3.3}{\nano\eV}& \textbf{$\sim$}\SIBold{200}{\nano\s}\\
$\gzwo$& \SIerr{9.2}{2}{\nano\eV}& \SIerrBold{71}{2}{\nano\s}\\
\bottomrule
    \end{tabular}
    \caption{\textbf{Comparison of extracted linewidths and characteristic times.} The extracted values are expressed in bold letters, while the converted values are non-bold. The conversion uses following proportionality: linewidth equals $\frac{\hbar}{\tau}$, with $\tau$ the expressed time and $\hbar$ the reduced Planck's constant~\cite{Jungwirth2016}. The expression IBM in the left column refers to the extrapolated zero temperature intercept in Fig.~\ref{fig: 2}(d), the expression $g^{(1)}(\tau)$ to the values as extracted from Fig.~\ref{fig: 3}, and $\tau_1$, $\tau_2$, $\tau_3$, $\tau^*_1$ to the values as in Fig.~\ref{fig: 4}(b) and (c). The expression $\gzwo$ refers to the measured decay times in Fig.~\ref{fig: 1}(b).
    }
    \label{tab: 2}
\end{table}

\section{4. Discussion}

Generally speaking, the utilized IBM allows us to describe the general shape of the photoluminescence spectra reasonably well at all relevant temperatures. However, above \SI{20}{\K}, the fitted width of the ZPL [cf. Fig.~\ref{fig: 2}(d)] appears to be impacted by the emergence of the high-energy tail of the photoluminescence due to phonon-absorption processes. These phonon processes may indeed influence the fundamental linewidth through additional dephasing channels, which would appear in the model as decreased phonon dephasing time. Moreover, we note that the exact lineshape of the SPE-related photoluminescence varies from cool-down to cool-down (cf. Supporting Information), as can be seen in detail by comparing the lowest extracted widths of the ZPL as in Fig.~\ref{fig: 2}(d) to the ones of Fig.~\ref{fig: 2}(f). The underlying reason is very likely a change of the specific electrostatic and dielectric environment of the SPE for each of the cool-downs ~\cite{Barthelmi2020}. For that reason, we presented the first-order autocorrelation $g^{(1)}(\tau)$ of the SPE-related photoluminescence after another temperature cycle (Fig.~\ref{fig: 3} and Supporting Information). Indeed, we find that the observed temporal coherence exhibits a process of about \SI{6}{\pico\s}, which is consistent with the lowest value of about \SI{110}{\micro\eV} as extracted for the width of the ZPL from all three cool-downs of this specific SPE (cf. Table~\ref{tab: 2} and Supporting Information). The presented GW-BSE calculations for a sulfur-vacancy show radiative times with a lower bound of a few 10s of picosecond within the energy range [dashed lines in Fig.~\ref{fig: 5}], where the experimentally investigated SPE exhibits the discussed ZPL. Consistently, the ZPL-linewidth of a SPE on a second sample shows values on the order \SI{70}{\micro\eV} ($\sim$ 9-\SI{10}{\pico\s}) (cf. square in \ref{fig: 2}(d) and Supporting Information). In order to estimate a lower bound of the ZPL based on the presented PL measurements, we examine the asymptotic behavior of the high-energy tail of the PL at the lowest temperature (cf. Supporting Information). We observe that below an experimental bath temperature of about 2 K, the asymptotic broadening at large detuning seems to be non-thermal and best described by a Lorentzian function with a HWHM of 30-60 µeV (cf. Supporting Information). We note that a resonant excitation scheme does not seem to be possible for the investigated SPEs in \ce{MoS2}, where one needs to utilize an excitation energy far above the SPE's emission energy resonant to an exciton-transition of \ce{MoS2} (cf. Supporting Information), to achieve the given signal to noise ratio~\cite{kumar_resonant_2016-1, Dietrich2018}.

As far as the time-resolved photoluminescence experiments are concerned, we observe a single-exponential decay time at low power and temperature [$\tau_1$ in Fig.~\ref{fig: 4}(b)] as well as at rather high bath temperatures above \SI{10}{\K} [$\tau^*_1$ in Fig.~\ref{fig: 4}(c)], which naively is in agreement with the assumed two-level system in the framework of the utilized IBM. However, the observed values of $\tau_1$ and $\tau^*_1$ of \SIerr{55}{1}{\nano\s} and \SIerr{200}{1}{\nano\s} (Table~\ref{tab: 2}) are up to four orders of magnitude longer than the value as suggested by the ZPL linewidth and the $g^{(1)}(\tau)$ (\SI{110}{\micro\eV}, $\sim$ \SI{6}{\pico\s}). We interpret the long apparent photoluminescence lifetimes to resemble necessary relaxation and phonon-scattering processes, such that the photo-excited SPE can recombine within the light cone. In this way, we also interpret the non-mono-exponential regime for higher power at low temperature [cf. gray shaded are in Fig.~\ref{fig: 4}(b)]. In particular, Muljarov and Zimmermann extended the IBM by including acoustic phonon-assisted transitions into energetically higher lying states, which gives rise to temperature-dependent broadening of the ZPL~\cite{muljarov_dephasing_2004}.  Additional asymmetric occupation of electron and hole states in the SPE can contribute to a polynomial decay, as discussed in ref. ~\cite{baer_luminescence_2006}. The dotted black line in the bottom panel of Fig.~\ref{fig: 4}(a) shows a numerical fit with a corresponding polynomial decay dependence (cf. Supporting Information).  According to the fit, we find a characteristic time of \SI{9}{\nano\s}, which is in the same order as the phenomenological time-scales $\tau_2$, and $\tau_3$.  At higher temperatures (above $\sim $\SI{10}{\K}), we observe again a mono-exponential decay time $\tau_1^*$ on the order of $\sim$\SI{200}{\nano\s} [white area Fig.~\ref{fig: 4}(c)], as it is consistent with earlier reports on SPEs in helium-ion treated \ce{MoS2}~\cite{Klein2019, KleinSigl2021}. We argue that at higher temperatures, phonon-absorption processes have to be taken into account. Since phonon-scattering processes are very likely to hinder the recombination of the SPE-related exciton within the light cone, they might explain the observation of an overall increase of the apparent luminescence lifetime and a decreasing intensity [Fig.~ \ref{fig: 1}(a) and Supporting Information]. Last but not least, we would like to mention that the timescales as detected by the $\gzwo$-measurements are consistent with the discussed apparent luminescence lifetimes in the nanosecond regime [cf. Figs.~\ref{fig: 1}(b) and (c), and Table~\ref{tab: 2}]. 

\section{5. Conclusions}

We investigated the zero-phonon line (ZPL) emission of single-photon emitters (SPEs) in helium-ion–treated monolayer \ce{MoS2}, which are currently understood in terms of sulfur-site vacancies, and established bounds on their homogeneous linewidths and coherence based on the presented photoluminescence spectra. A line shape analysis describing coupling of the SPE to acoustic phonons within the independent boson model (IBM) yields an upper bound of the ZPL linewidth of $\sim$\SI{110}{\micro\eV} below \SI{20}{\K} and a lower bound of $\sim$\SI{30}{\micro\eV} below \SI{2}{\K}. Future resonant excitation schemes might give lower values. Ab initio GW-BSE calculations for sulfur-vacancies predict a lower-bound for the radiative lifetimes of tens of picoseconds near the ZPL energy, supporting the interpretation of the experimental results. First-order coherence measurements are consistent with the ZPL values, while the phonon sideband exhibits a sub-picosecond coherence. Moreover, we observe Debye–Waller factors up to $\sim$30-\SI{40}{\%} depending on cool-down and excitation power. The apparent luminescence lifetimes are in the regime from several to hundreds of nanoseconds, suggesting that relaxation- and phonon-scattering processes dominate the recombination via the light cone.

\section*{Acknowledgments}
\noindent We gratefully acknowledge the German Science Foundation (DFG) for financial support via the clusters of excellence MCQST (EXS-2111) and e-conversion (EXS-2089), and the priority program 2244 (2DMP) via HO3324/13-2 and KA 5418/1-2 as well as the Munich Quantum Valley (K6). K.W. and T.T. acknowledge support from the JSPS KAKENHI (Grant Numbers 21H05233 and 23H02052), the CREST (JPMJCR24A5), JST and World Premier International Research Center Initiative (WPI), MEXT, Japan. T.A. acknowledges support from the Azrieli Graduate Fellows Program. S.R.A. acknowledges support from a European Research Council (ERC) Starting Grant (No. 101041159). Computational resources were provided by the ChemFarm local cluster at the Weizmann Institute of Science and the Oak Ridge Leadership Computing Facility through the Innovative and Novel Computational Impact on Theory and Experiment (INCITE) program, which is a DOE Office of Science User Facility supported under Contract No. DE-AC05-00OR22725. 

\section*{Author contributions}
\noindent KB, MT, LS, AM, TD, SG, VZ, CB performed the experiments. TA, MF, ML, SRA performed the theoretical calculations. AWH and SRA perceived the project. TT and KW supplied the high-quality hBN. All authors contributed to the writing, read and approved the manuscript.

\subsection{Competing interests}
\noindent All authors declare no financial or non-financial competing interests. 

\subsection{Data availability}
 
\noindent The data analyzed in the current study is available from the authors upon reasonable request.

\bibliography{preambles/bibliography}

\end{document}


\newcommand{\wsi}{Walter Schottky Institut and Physics Department, Technical University of Munich, Am Coulombwall 4a, 85748 Garching, Germany.}
\newcommand{\mcqst}{MCQST, Schellingstrasse 4, 80799 München, Germany.}
\newcommand{\kth}{KTH Royal Institute of Technology, Department of Applied Physics, Albanova University Centre, Roslagstullsbacken 21, 106 91 Stockholm, Sweden}
\newcommand{\mich}{Department of Electrical Engineering and Computer Science, University of Michigan, 1301 Beal Avenue, Ann Arbor, Michigan 48109, United States}
\newcommand{\weizmann}{Department of Molecular Chemistry and Materials Science, Weizmann Institute of Science, 7610001 Rehovot, Israel.}
\newcommand{\Duisburg}{Faculty of Physics, University of Duisburg-Essen, Germany.}
\newcommand{\takashi}{Research Center for Materials Nanoarchitectonics, National Institute for Materials Science,  1-1 Namiki, Tsukuba 305-0044, Japan}
\newcommand{\watanabe}{Research Center for Electronic and Optical Materials, National Institute for Materials Science, 1-1 Namiki, Tsukuba 305-0044, Japan}

\author{Katja Barthelmi}
\affiliation{\wsi}
\affiliation{\mcqst}

\author{Tomer Amit}
\affiliation{\weizmann}

\author{Mirco Troue}
\affiliation{\wsi}
\affiliation{\mcqst}

\author{Lukas Sigl}
\affiliation{\wsi}

\author{Alexander Musta}
\affiliation{\wsi}

\author{Tim Duka}
\affiliation{\wsi}

\author{Samuel Gyger}
\affiliation{\kth}

\author{Val Zwiller}
\affiliation{\kth}

\author{Matthias Florian}
\affiliation{\mich}

\author{Michael Lorke}
\affiliation{\Duisburg}

\author{Takashi Taniguchi }
\affiliation{\takashi}

\author{Kenji Watanabe}
\affiliation{\watanabe}

\author{Christoph Kastl}
\affiliation{\wsi}
\affiliation{\mcqst}

\author{Jonathan Finley}
\affiliation{\wsi}
\affiliation{\mcqst}

\author{Sivan Refaely-Abramson}
\affiliation{\weizmann}

\author{Alexander Holleitner}
\email[]{holleitner@wsi.tum.de}
\affiliation{\wsi}
\affiliation{\mcqst}

\title{Supporting Information for: Zero-phonon line emission of single photon emitters in helium-ion treated \texorpdfstring{\ce{MoS2}}{MoS2}}
\date{December 1, 2025}
\maketitle
\tableofcontents
\newpage

\section{Further photoluminescence characteristics of the investigated SPE on sample 1}
\begin{figure*}[h]
\centering
\includegraphics[scale=1]{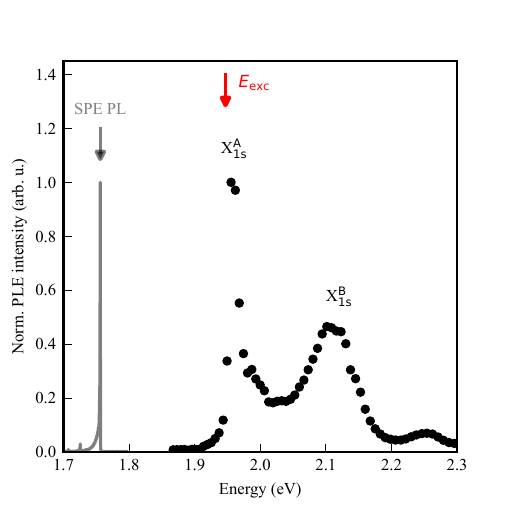}
\caption{\textbf{Photoluminescence excitation (PLE) of the SPE on sample 1}.  All measurements as presented in the main manuscript are performed with an excitation energy of \Eexc\,=\,\SI{1.943}{\eV}, which is close to the resonance of the neutral exciton \neutX\, of \ce{MoS2}. To demonstrate this, we perform PLE experiments on the SPE on sample 1 (cf. PL spectrum displayed in grey). (\Pexc\,=\,\SI{2.5}{\micro\W}). The PLE (black data) clearly shows the neutral $A$ and $B$ excitons of \ce{MoS2} at \SI{1.947}{eV} (X$^\text{A}_\text{1S}$) \cite{Wierzbowski2017} and X$^\text{B}_\text{1S}$ at an energy offset of $\sim$\SI{150}{\milli\eV} to the $A$ exciton \cite{Vaquero2020} (\Pexc\,=\,\SI{2.5}{\micro\W}.)}
\label{figSI: ple}
\end{figure*}
\begin{figure*}[h]
\centering
\includegraphics[scale=1]{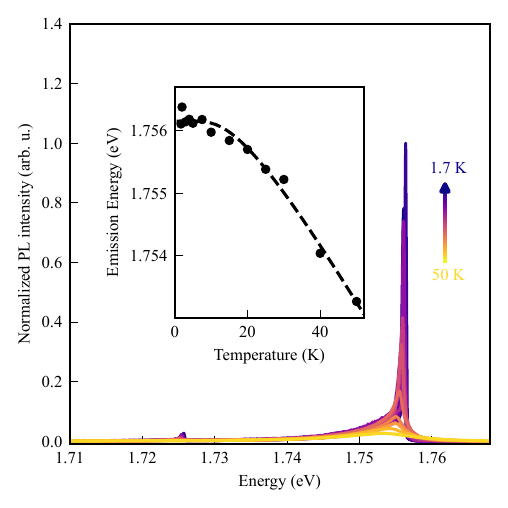}
\caption{\textbf{Temperature dependent photoluminescence spectra of the investigated SPE on sample 1.} \textbf{(a)} Photoluminescence spectra as in Fig.~2(b) of the main manuscript without normalization. All data are measured with \Eexc\,$ = \SI{1.94}{eV}$ and \Pexc\,$=\SI{500}{\nano\W}$. Inset shows the temperature dependent shift of the energy of the zero-phonon line and the data is fitted by a phenomenological energy shift (dashed line and \cite{ODonnell1991}). From the fit we get the energy at \SI{0}{\K} of $E_\text{ZPL}(0)=\SI{1.7562}{\eV}$, the Huang-Rhys factor (HRF) $S$ of \SIerr{0.6}{1}{} (at \SI{0}{\K}~\cite{Jungwirth2016}) and an average phonon energy $\hbar \omega$ of \SI{4.4}{\milli\eV}. Note that the temperature dependent shift includes both a contribution from a polaron shift and a band gap and exciton binding energy renormalization due to the interplay of anharmonic lattice expansion and changes in screening.}
\label{figSI: polaron}
\end{figure*}

\begin{figure*}[h]
\centering
\includegraphics[scale=0.97]{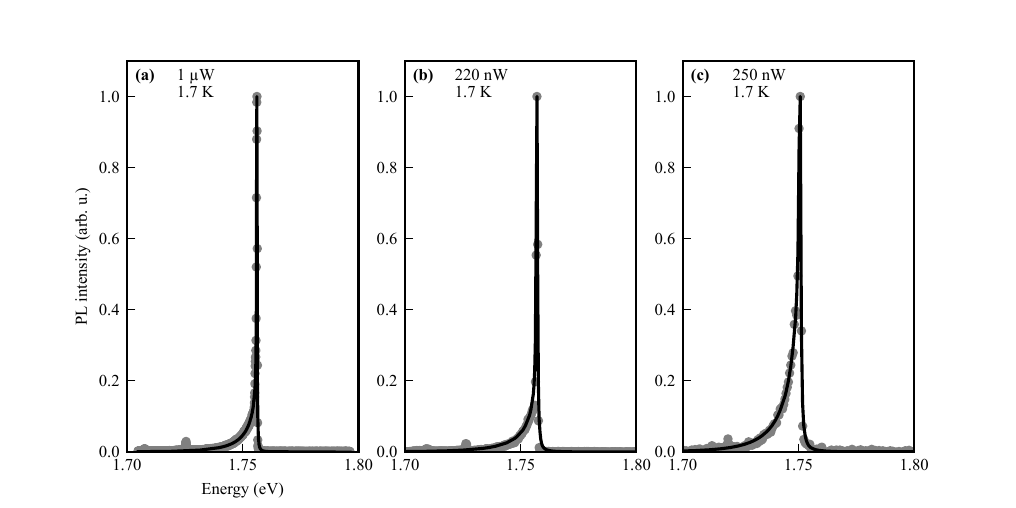}
\caption{\textbf{Comparison of photoluminescence spectra of the investigated SPE on sample 1 for three cooldowns.} \textbf{(a)} Spectrum as in Fig.~2(d). The IBM fit results in an HRF of 1.37 and a homogeneous broadening of \SIerr{118}{4}{\micro\eV}. (b) Spectrum as in Fig.~2(e). The IBM fit results in an HRF of 1.03 and a homogeneous broadening of \SIerr{300}{30}{\micro\eV}. (c) Spectrum before first-order correlation measurement as in Fig.~3. The IBM fit results in a HRF of 1.61 and a homogeneous broadening of \SIerr{370}{10}{\micro\eV}.}
\label{figSI: coherence}
\end{figure*}

\begin{figure*}[h]
\centering
\includegraphics[scale=0.97]{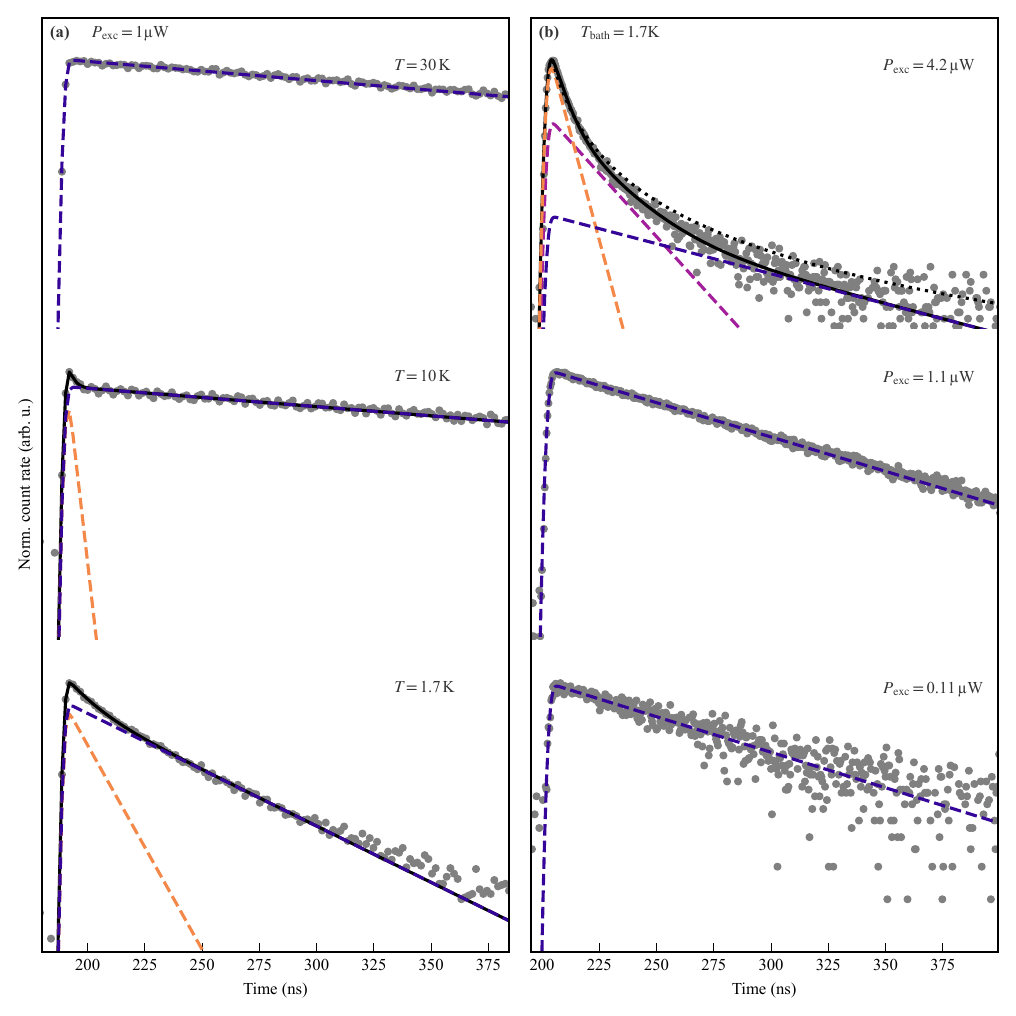}
\caption{\textbf{Time-resolved luminescence of the SPE on sample 1.} (a) Exemplary original time-resolved photoluminescence traces with fits (dashed lines) for the results as presented in Fig. 4(c) of the main manuscript. (b) Identical data as in Fig. 4(a) of the main manuscript with exponential fits in blue, purple, and orange. Black line
at top panel is sum of blue, purple, and orange dashed lines. Black
dotted line is a polynomial fit (cf. Supporting Section on Analytical model of time-resolved photoluminescence).}
\label{figSI: decay}
\end{figure*}

\clearpage

\section{Photoluminescence spectrum of the investigated SPE on sample 2}
\begin{figure*}[h]
\centering
\includegraphics[scale=0.97]{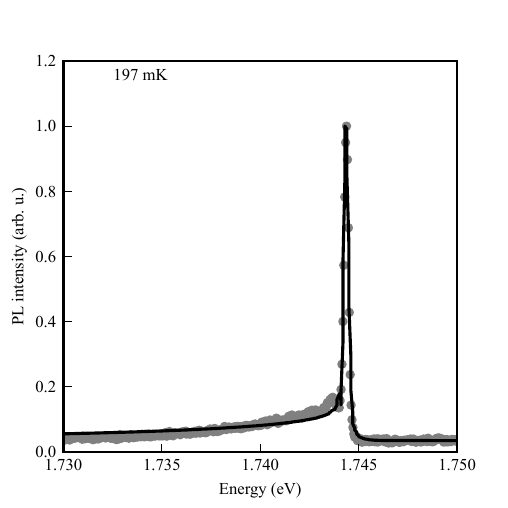}
\caption{\textbf{Photoluminescence spectrum of sample 2 at \SI{197}{\milli\K.}} The IBM fit results in a HRF of 1.41 and a homogeneous broadening of \SI{70}{\micro\eV}.}
\label{figSI: coherence}
\end{figure*}

\section{Temporal coherence length}
\begin{figure*}[h]
\centering
\includegraphics[scale=0.97]{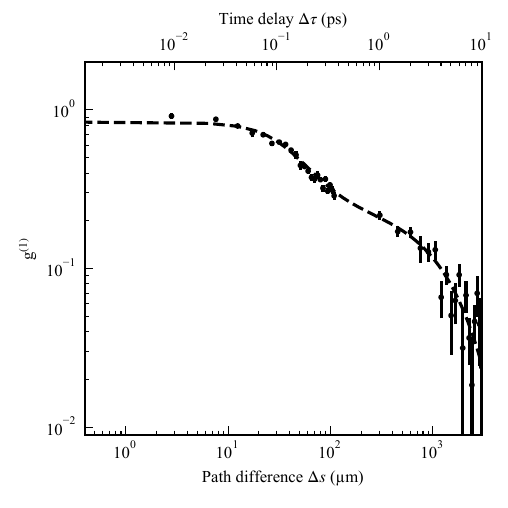}
\caption{\textbf{First-order auto-correlation measurement of
the SPE on logarithmic scale.} Data as in Fig. 3 of main manuscript in a logarithmic presentation to demonstrate the quality of the fit (dashed line).}
\label{figSI: coherence}
\end{figure*}

\section{Lineshape analysis of the high-energy tail of the ZPL}
\begin{figure*}[h]
\centering
\includegraphics[scale=1]{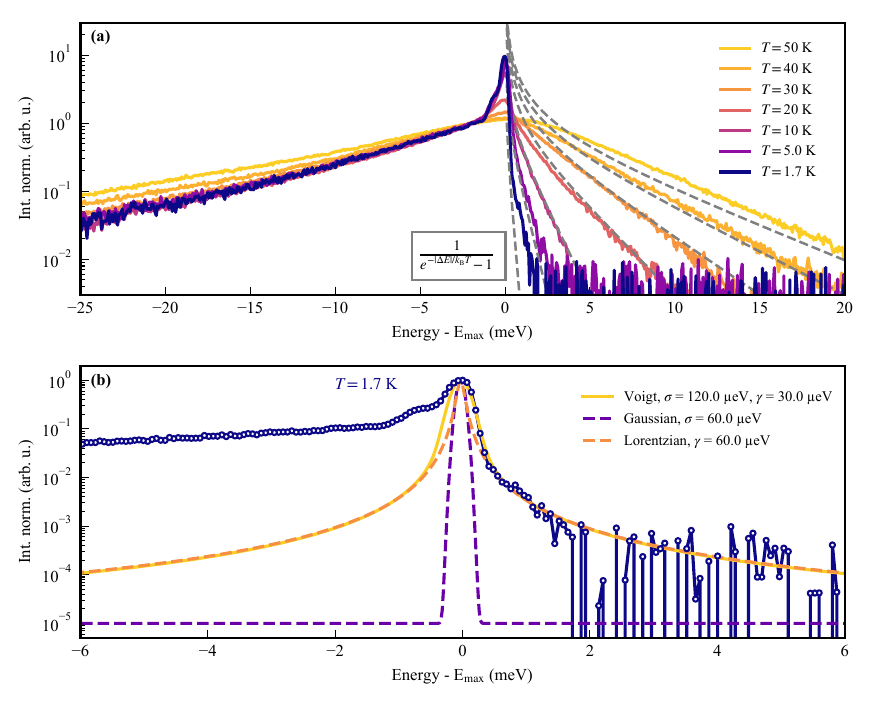}
\caption{\textbf{Asymptotic behavior of the temperature dependent PL spectra}.  \textbf{(a)} Temperature dependent PL spectra on a logarithmic scale. The data are normalized to the intensity of the onset of the low-energy phonon tail (amplitude after the ZPL). The energies are referenced to the energy of the peak maximum $E_\text{max}$. The high energy tail (phonon absorption) at $E > E_\text{max}$ shows an exponential behavior consistent with a thermal broadening described by a Bose-Einstein distribution at the respective temperature (gray dashed lines). Importantly, we find that below \SI{2}{K}, the asymptotic behavior of the lineshape (large positive $E-E_\text{max}$) is not governed by the thermal phonon population, but by the intrisinc lineshape of the ZPL. \textbf{(b)} Analysis of the ZPL at \SI{1.7}{K}. The lineshape of the ZPL reveals homogeneous broadening described by a Lorentzian (orange dashed line) with \SI{60}{\micro eV} broadening (HWHM). A single Gaussian (dashed purple line) is clearly insufficient to describe the lineshape. Alternatively, a Voigt function with \SI{120}{\micro eV} Gaussian boradening and \SI{30}{\micro eV} Lorentzian broadening (yellow solid line) provides a good model for the lineshape. This places the lower bound of the homogeneous broadening within \SI{30}{\micro eV} - \SI{60}{\micro eV}.}
\label{figSI:log}
\end{figure*}

\section{Analytical model of time-resolved photoluminescence}

The fit as presented as a black dotted curve in the bottom panel of Fig. 4(a) of the main manuscript is based on a quantum-optical many-body treatment of correlations within the cluster expansion framework \cite{baer_luminescence_2006}.
Here, we examine whether the occupation of an electronic defect state is necessarily accompanied by the presence of a hole in the corresponding valence-band defect state. In general, the temporal evolution of the carrier populations in these states due to PL can be expressed as
\begin{equation}
\dot{f}^{e,h}(t)=-\frac{f^e(t)f^h(t)+\text{correlations}}{\tau}.
\end{equation}
In the case of perfect correlations—for example, under resonant excitation of a two-level system—an electron is \emph{always} accompanied by a hole. In this limit, the numerator reduces to the electron or hole occupation, recovering the well-known expression
\begin{equation}
\dot{f}^{e,h}(t)=-\frac{f^{e,h}(t)}{\tau},
\end{equation}
which leads to a purely exponential decay.

In the present case, however, the electronic defect state lies deep within the band gap, whereas the hole state is resonant with valence-band states. We therefore consider the opposite limit: carriers can scatter freely into and out of the valence-band defect state, effectively destroying correlations. This yields to following expression,
\begin{equation}
\dot{f}^{e,h}(t)=-\frac{f^e(t)f^h(t)}{\tau},
\end{equation}
and the corresponding time-dependent PL intensity is given by
\begin{equation}
I(t)=-\frac{f^e(t)f^h(t)}{\tau}.
\end{equation}
This expression was used for the non-exponential fit as shown as a dotted curve in Fig.~4(a) of the main text and supporting Fig. S4(b).

\section{Computational details of the GW and BSE calculations}

First principles calculations are employed within many-body perturbation theory, where density functional theory (DFT) serves as a starting point~\cite{kohn1965self}. We first compute the electronic wavefunctions and Kohn-Sham energies from DFT with the Perdew-Burke-Ernzerhof (PBE) exchange-correlation functional~\cite{perdew1996generalized} within the Quantum Espresso package~\cite{giannozzi2009quantum, giannozzi2017advanced}, including spin-orbit coupling and explicit spinor wavefunctions. The computed system contains a relaxed $5\times5\times1$ supercell of monolayer MoS$_2$ with a single sulphur vacancy per supercell, and 15\textrm{\AA} separation between periodic layers in the out-of-plane direction. The calculations were done with a Bloch plane-wave basis-set, using norm-conserving pseudopotentials from PseudoDojo~\cite{van2018pseudodojo} and with a 75~Ry energy cutoff. The self-consistent calculations of the electron density were done on a $6\times6\times1$ uniform k-grid.
The quasiparticle bandstructure and optical properties were computed within the GW-BSE approximation using the BerkeleyGW software~\cite{deslippe2012berkeleygw}. Quasiparticle energy corrections were calculated within $G_0W_0$ and the generalized plasmon-pole approximation~\cite{hybertsen1986electron}, on a uniform $3\times3\times1$ k-point grid and with 3998 spinor bands and 25~Ry cutoff of the screening function. We then expand our sampling through the Nonuniform Neck Subsampling (NNS) scheme~\cite{jordana2017nonuniform} to capture the behavior of the electron wavefunctions accurately throughout the Brillouin zone.
Optical properties were computed by solving the Bethe-Salpeter equation (BSE)~\cite{rohlfing2000electron, *rohlfing1998electron}. The electron-hole interaction kernel was computed using a dielectric matrix with 1798 bands on a $6\times6\times1$ k-point grid with 5~Ry screening cutoff. Exciton wavefunctions and energies were evaluated by interpolation of the interaction kernel onto a $18\times18\times1$ grid and diagonalization of the BSE Hamiltonian. We then calculated the radiative decay time of each excitonic state as detailed in the main manuscript.

\bibliography{preambles/bibliography}